\pdfoutput=1
\documentclass[10pt,twocolumn,letterpaper]{article}

\usepackage{cvpr}
\usepackage{times}
\usepackage{epsfig}

\usepackage[latin1]{inputenc} 
\usepackage[T1]{fontenc}    

\usepackage[numbers,compress]{natbib}
\usepackage{footmisc} 
\usepackage[dvipsnames]{xcolor}
\usepackage{footnote} 
\usepackage[pagebackref=true,breaklinks=true,letterpaper=true,colorlinks,bookmarks=false]{hyperref}
\usepackage{url}    

\usepackage{booktabs}       
\usepackage{multirow}
\usepackage{tabularx}
\usepackage{color, colortbl}

\usepackage{amsmath,amsfonts,amsthm,amssymb}
\usepackage{stmaryrd} 
\usepackage{mathtools}
\usepackage[separate-uncertainty=true]{siunitx}
\usepackage{nicefrac}       
\usepackage{microtype}      
\usepackage{mathrsfs}

\usepackage{subcaption}
\usepackage{graphicx}
\usepackage[export]{adjustbox}
\usepackage{rotating}
\graphicspath{{./figures/}{./figure1/}}

\usepackage{pifont}
\makeatletter
\@namedef{ver@everyshi.sty}{}
\makeatother
\usepackage{tikz}
\usepackage{xifthen}		
\usepackage{pgfplots}	
\usepgflibrary{patterns}
\usetikzlibrary{%
  arrows,%
  calc,%
  fit,%
  patterns,%
  plotmarks,%
  shapes.geometric,%
 shapes.misc,%
  shapes.symbols,%
  shapes.arrows,%
  shapes.callouts,%
  backgrounds,%
  chains,%
  topaths,%
  trees,%
  petri,%
  matrix,%
  folding,%
  fadings,%
  through,%
  positioning,%
  scopes,%
  decorations.fractals,%
  decorations.shapes,%
  decorations.text,%
  decorations.pathmorphing,%
  decorations.pathreplacing,%
  decorations.footprints,%
  decorations.markings,%
  shadows,%
  bayesnet,%
  quotes,%
  arrows.meta,%
  graphs%
}

\newcommand{\bfcaption}[2]{\caption[#1]{#1. #2}}     

\newcommand{\fett}[1]{\mathbf{#1}}

 \cvprfinalcopy 

\ifcvprfinal\fi
\begin{document}

\title{DISCo: Deep learning, Instance Segmentation, and Correlations for cell segmentation in calcium imaging}

\author{Elke Kirschbaum \and Alberto Bailoni \and Fred A. Hamprecht\\
Interdisciplinary Center for Scientific Computing (IWR)\\
Heidelberg University, Heidelberg, Germany\\
{\tt\small \{elke.kirschbaum,alberto.bailoni,fred.hamprecht\}@iwr.uni-heidelberg.de}
}

\maketitle

\begin{abstract}
Calcium imaging is one of the most important tools in neurophysiology as it enables the observation of neuronal activity for hundreds of cells in parallel and at single-cell resolution. In order to use the data gained with calcium imaging, it is necessary to extract individual cells and their activity from the recordings. We present DISCo, a novel approach for the cell segmentation in calcium imaging videos. We use temporal information from the recordings in a computationally efficient way by computing correlations between pixels and combine it with shape-based information to identify active as well as non-active cells. We first learn to predict whether two pixels belong to the same cell; this information is summarized in an undirected, edge-weighted grid graph which we then partition. In so doing, we approximately solve the NP-hard correlation clustering problem with a recently proposed greedy algorithm. Evaluating our method on the Neurofinder public benchmark shows that DISCo outperforms all existing models trained on these datasets. 
\end{abstract}


\section{Introduction}
Calcium imaging is a microscopy technique that allows the observation of the activity of large neuronal populations at single-cell resolution~\citep{Denk73, Helmchen2005DeepTT, flusberg_high-speed_2008}. This makes it one of the most important tools in neurophysiology since it enables the study of the formation and interaction of neuronal networks in the brain. The data recorded with calcium imaging is a sequence of images - the calcium imaging video - that shows multiple cells, that have a fixed position over the whole length of the video but varying luminosity depending on their activity. 
The extraction of the individual cell locations from the calcium imaging videos is a fundamental but yet unsolved problem in the analysis of this data~\citep{lemonadeICLR}.

In order to encourage the development of new tools for the cell segmentaiton in calcium imaging videos and to enable a meaningful comparison of different approaches, the Neurofinder public benchmark~\citep{Neurofinder} was initiated. The Neurofinder challenge consists of 19 calcium imaging videos with ground truth cell annotations for training, and of nine test videos with undisclosed ground truth. Both, training and test set, can be clustered into five dataset series (named 00, 01, 02, 03, and 04) which were recorded under different conditions and differ also in labeling technique and whether the ground truth annotations contain mainly active, inactive or both kinds of cells. Details on the five groups of datasets can be found e.g.\ in~\citet{spaen_2017}. 

In this paper, we present \emph{DISCo}, a novel approach using \emph{Deep learning, Instance Segmentation, and Correlations} for the cell segmentation in calcium imaging videos. DISCo combines the advantages of a deep learning model with a state-of-the-art instance segmentation algorithm, allowing the direct extraction of cell instances. Additionally, we use temporal context from the calcium imaging videos by computing segment-wise correlations between pixels. This temporal information is combined with shape-based information, which is a huge advantage of DISCo compared to methods that solely rely on the one or the other. This enables us to achieve a very good overall performance using only a single model on all Neurofinder datasets. Moreover, when training individual networks on the five dataset series (submission called DISCos), we are able to outperform all other methods trained on the Neurofinder datasets.


\section{Related Work}\label{sec:bg_cellex}
For the extraction of cells from calcium imaging data most existing algorithms are based on non-negative matrix factorization (NMF)~\citep{mukamel_09_automated,pnevmatikakis_13_sparse,pnevmatikakis_13_rank,pnevmatikakis_structured_2014,andilla2014sparse,maruyama_14_detecting,PNEVMATIKAKISdemixing,friedrich2017fast,NIPS2017_inan,giovannucci2017onacid,pnevmatikakis2018,giovannucci2019caiman}, 
clustering~\citep{Kaifosh2014SIMAPS,spaen_2017}, 
dictionary learning~\citep{diego_13_automated,diego_13_learning,pachitariu_13_extracting,petersen2018scalpel}, and 
deep learning~\citep{CACNN,Klibisz17,Ca3dCNN}. 
In the Neurofinder leaderboard\footnote{Leaderboard of the Neurofinder challenge at \url{http://neurofinder.codeneuro.org}. Accessed: 2019-11-15. We do not discuss the results of the submissions Mask R-CNN and human-label since we have no information on the used models and training procedures.} the currently top scoring methods are STNeuroNet and 3dCNN~\citep{Ca3dCNN}\footnote{According to private correspondence with \citeauthor{Ca3dCNN} 3dCNN is a developmental stage of STNeuroNet~\citep{Ca3dCNN}.}, followed by HNCcorr~\citep{spaen_2017} combined with Conv2D~\citep{conv2d}, UNet2DS~\citep{Klibisz17}, as well as Sourcery and Suite2P~\citep{pachitariu2017suite2p} together with Donuts~\citep{pachitariu_13_extracting}.

UNet2DS~\citep{Klibisz17} and Conv2D~\citep{conv2d} use deep learning models with so-called {\it summary images} as input. These summary images contain for each pixel the mean projection over time, which means that all temporal information of the calcium imaging videos is lost. As a consequence, these approaches are not competitive on datasets which contain many active neurons, like the dataset series 01, 02 and 04 of the Neurofinder challenge. In contrast to this, the method HNCcorr~\citep{spaen_2017} is able to detect the active cells in the dataset series 01, 02 and 04 fairly well, while it performs rather poorly on the other datasets. The reason for this is that HNCcorr uses a clustering algorithm based on the distance of pixels in {\it correlation space}. In this correlation space pixels from cells with a changing signal should be well separated from background pixels, but pixels from cells with weak or constant activity will not be distinguishable from background. In order to overcome this problem and to achieve competitive average F1-scores in the Neurofinder challenge, \citet{spaen_2017} combined HNCcorr and Conv2D by using the first for the dataset series 01, 02 and 04 and the latter for the series 00 and 03. The same holds true for the NMF-based methods Sourcery and Suite2P~\citep{pachitariu2017suite2p} which need to be complemented by the shape-based algorithm Donuts~\citep{pachitariu_13_extracting} in order to achieve decent average F1-scores over all dataset series. 

In contrast to this, the deep learning models STNeuroNet~\citep{Ca3dCNN} and its developmental stage 3dCNN are able to achieve good F1-scores on all test datasets with a single method using a 3D convolutional neural network (CNN) on the calcium imaging video. Since such models can become computationally very costly, especially for videos consisting of several thousand frames like the ones in the Neurofinder challenge, the models are only run on short temporal batches of the video and to gain the cell locations for the whole video, the outputs from the different batches have to be merged in the post-processing. Moreover, like all leading methods using deep learning, STNeuroNet and 3dCNN only provide a foreground-background prediction and also need post-processing to extract individual cell instances. 
STNeuroNet was trained with additional data from the Allen Brain Observatory (ABO) dataset~\citep{abo} and with manually refined ground truth. In contrast to this, 3dCNN uses only the datasets and ground truth provided in the Neurofinder challenge. Though the 3dCNN submission consists of a single method, it uses separately trained networks for each of the five Neurofinder dataset series.


\section{Method}\label{sec:disco_method}
Broadly speaking, DISCo extracts temporal information from the calcium imaging videos by computing segment-wise correlations between pixels. This temporal information is combined with shape-based information from a summary image and transformed to affinities between pixels by a deep neural network. Finally, the affinities are used by a state-of-the-art instance segmentation algorithm to extract and separate individual cells. 

More specifically, DISCo starts by splitting the video temporally into segments on which the correlations between pixels are computed as described in section~\ref{sec:disco_corrs}. The idea of a segment-wise computation of the correlation coefficient was proposed by \citet{nikolic2012scaled} and has already been successfully used in the analysis of neuronal activity data~\citep{nikolic2012scaled,folias2013synchronisation,dolean2017scaled}. The benefits of the segment-wise correlations are: an improvement of the signal-to-noise ratio (SNR); and more fine-grained information about the temporal dynamics of the pixels.

Computing the correlations on multiple segments means that the temporal dimension of the video tensor is reduced, but is not completely removed. Hence, we use a small 3D CNN to aggregate the information from the different segments. In addition, a summary image is computed by taking for each pixel the mean projection over the whole video. The summary image is combined with the aggregated information from the segment-wise correlations to provide the second network with temporal and shape-based information. The second network maps this input to affinities between pixels in a highly non-linear fashion. The details of the two networks and how they are trained are given in section~\ref{sec:method_dl}. 

In the final step, an undirected, edge-weighted graph is constructed from the predicted affinities and the individual cells are directly extracted and separated by partitioning this signed graph. In addition to the pixel-wise affinities, the neural network also provides a foreground-background prediction which is used in the instance segmentation algorithm to directly exclude background pixels from the graph before the clustering, reducing false merges of cells and background. The details of the instance segmentation algorithm are described in section~\ref{sec:disco_gasp}. 

\begin{figure*}[t]
\centering
\adjustbox{max width=\textwidth}{
\includegraphics[width=\textwidth]{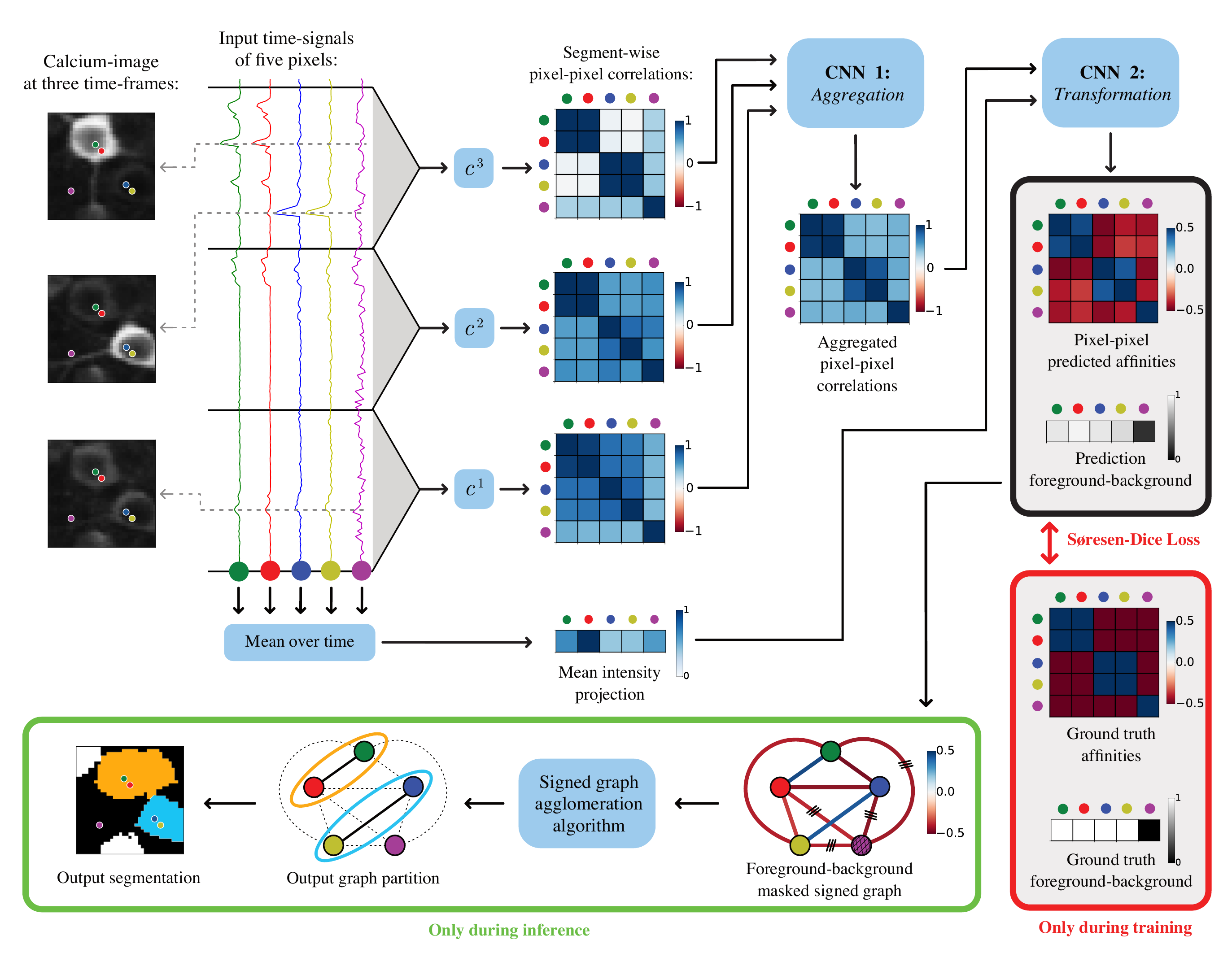}
}
\bfcaption{DISCo workflow}{For five exemplary pixel signals from the Neurofinder training set we show how they are processed by DISCo. First, the signals are split into temporal segments and segment-wise correlations between the pixels are computed. Next, the information from the segments is aggregated by a first CNN. These aggregated correlations are complemented by a mean intensity projection over time for each pixel. This combination of temporal and spatial information is processed by a second CNN which outputs affinities between pixels and a foreground-background prediction. Both CNNs are trained end-to-end. The predicted affinities are used by a graph partitioning algorithm to gain the final instance segmentation. \label{fig:disco_workflow}}
\end{figure*}

The complete model is also summarized in figure~\ref{fig:disco_workflow} and a PyTorch implementation of the proposed method is available on GitHub.\footnote{\url{https://github.com/EKirschbaum/DISCo}}

\begin{figure}[t]
\centering
\includegraphics[width=0.45\textwidth]{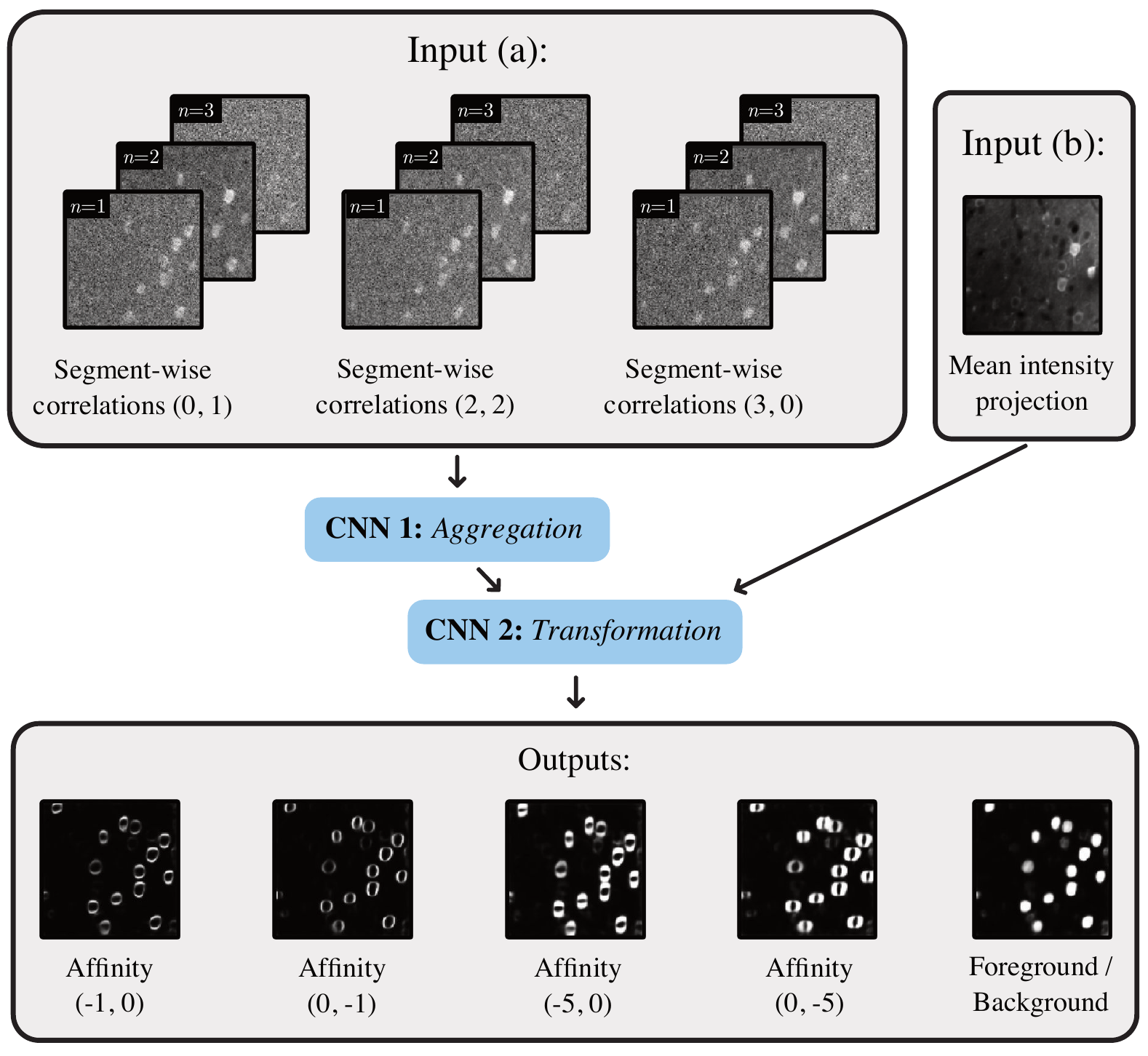}
\bfcaption{DISCo inputs and outputs}{We show the inputs and outputs of the (trained) deep learning model of DISCo for a crop of the Neurofinder training set. Segment-wise correlations (input (a)) are given as input to the first CNN. In this example, we show three segments ($N=3$) and the correlations to three exemplary neighbors ($C=3$). We show the offset to the considered neighbor in brackets. 
The mean intensity projection (input (b))  is then concatenated to the output of the first CNN and the combination is given as input to the second CNN model. The final output consists of a foreground-background prediction and pixel affinities in different directions, where the offsets are again given in brackets. \label{fig:disco_model}}
\end{figure}

\subsection{Temporal Information from Correlations}\label{sec:disco_corrs}
Since the fluorescence dynamics of cells and those of background pixels differ drastically, using the temporal context from calcium imaging videos is a huge benefit for the detection of cells. Moreover, without temporal information it is difficult to separate touching or overlapping cells correctly. 
For this reason, we use the temporal information from the calcium imaging videos in form of correlations. 

Consider a video $\fett{X}\in\mathbb{R}^{T\times P \times Q}$ with $T$ time frames and $P\times Q$ pixels. We define the vector $\fett{x}_{pq}$ to be the signal of pixel $(p,q)$ of length $T$ with $\fett{x}_{pq}(t)=\fett{X}_{tpq}$. For two pixels $(p,q)$ and $(p',q')$ the Pearson correlation coefficient~\citep{pearson1895notes} between their signals is given by
\begin{align}\label{eq:disco_pearson}
c(\fett{x}_{pq},\fett{x}_{p'q'}) &= \frac{\left<\fett{x}_{pq}-\bar{\fett{x}}_{pq},\fett{x}_{p'q'}-\bar{\fett{x}}_{p'q'} \right>}{\| \fett{x}_{pq}-\bar{\fett{x}}_{pq}\|_2 \cdot \|\fett{x}_{p'q'}-\bar{\fett{x}}_{p'q'}\|_2} \quad ,
\end{align}
where $\bar{\fett{x}}_{\cdot\cdot}$ denotes the mean of the signal $\fett{x}_{\cdot\cdot}$ over time and $\left<\cdot,\cdot\right>$ is the dot product. The Pearson correlation coefficient measures the linear correlation between the two signals and is $1$ for perfectly correlated signals, $0$ for non-correlated signals and $-1$ for anti-correlated signals. In theory it might seem beneficial to use other measures that can also take into account non-linear associations between signals, like e.g.\ the distance correlation~\citep{szekely2007measuring}. In practice, however, we found the Pearson correlation to be the better choice, since it can be computed faster and very efficiently even for large images and long time series. This allows us to compute the correlations online during the network training, which enables a broader range of data augmentation steps. 

Instead of computing the correlations between two pixels over the whole temporal extent of the video, we first split the video into $N=10$ segments and then compute the correlations segment-wise. The correlation between two pixels during the $n$-th segment is then given by
\begin{align}\label{eq:disco_pearson_segs}
c^n(\fett{x}_{pq},\fett{x}_{p'q'}) = c(\fett{x}^n_{pq},\fett{x}^n_{p'q'})
\end{align}
with $c(\cdot,\cdot)$ as defined in equation~\eqref{eq:disco_pearson} and $\fett{x}^n_{\cdot\cdot}$ being the $n$-th segment of the time series $\fett{x}_{\cdot\cdot}$. The segment-wise computation of correlation coefficients was proposed by \citet{nikolic2012scaled}, who defined the {\it scaled correlation} $c_s$ to remove correlation components slower than those defined by the used scale $s=T/N$, with
\begin{align}
c_s(\fett{x}_{pq},\fett{x}_{p'q'}) = \frac1N \sum_{n=1}^N c^n(\fett{x}_{pq},\fett{x}_{p'q'}) \quad .
\end{align}
Note that although Pearson correlation coefficients are a measure of linear association, their computation is non-linear and hence $c_s\neq c$ for $N>1$. 

In contrast to \citet{nikolic2012scaled}, we not only take the mean of the segment-wise correlations, but use a CNN to aggregate the information of the different segments. As shown in section~\ref{sec:disco_results} this provides even more information for the transformation network and results in a better segmentation. 

\def\segcorrs{1/0.92, 2/0.61,3/0.45,4/0.70,5/0.78,6/0.45,7/0.87,8/0.43,9/0.91,10/0.72}
\begin{figure*}[t]
\centering
\adjustbox{max width = .7\textwidth}{
\begin{tikzpicture}

\node[inner sep=0] (xpq) at (-1,-4.5) {$\fett{x}_{pq}$};
\node[inner sep=0] (xpqbar) at (-1,-7) {$\fett{x}_{p'q'}$};

\node[inner sep=0] (traces) at (7,-5) {\includegraphics[width=.875\textwidth]{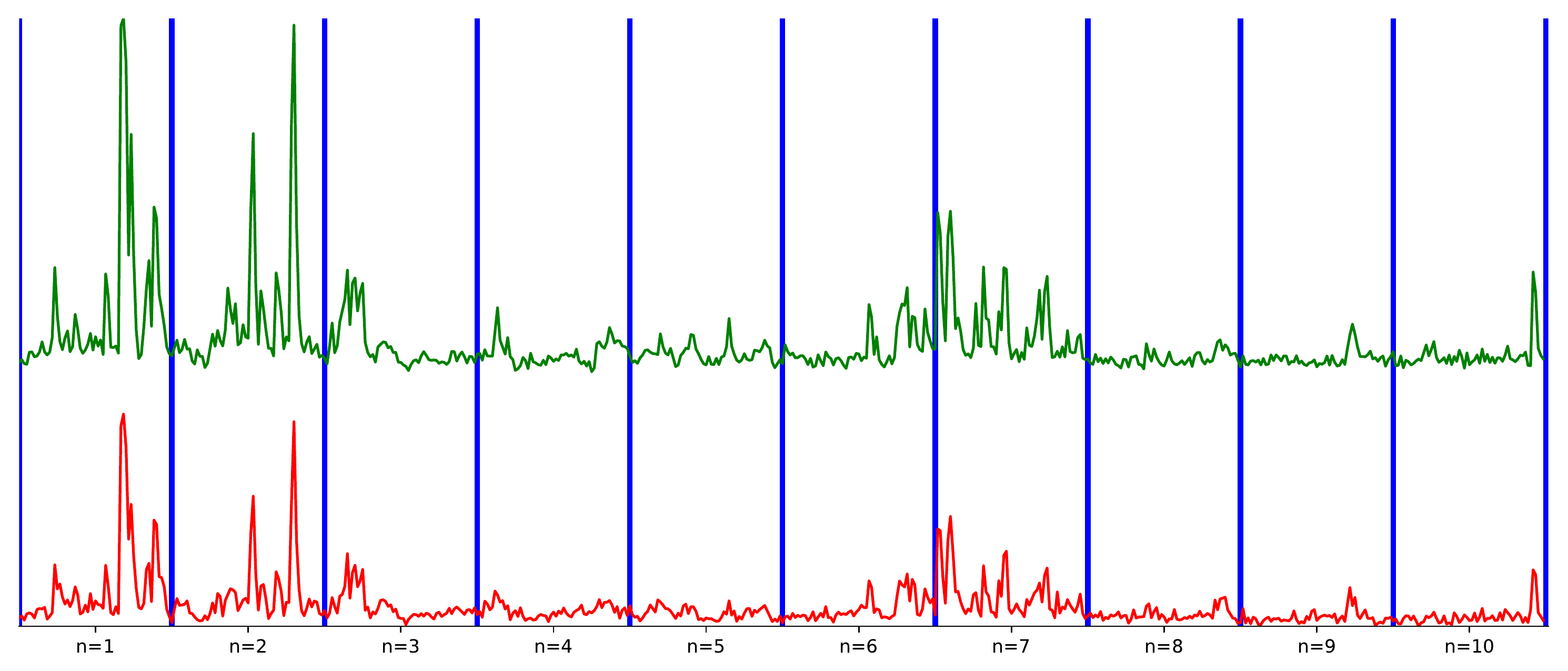}};

\node[inner sep=0] (cn) at (-1,-9) {$c^n:$};
\foreach \x/\y in \segcorrs
	\node[inner sep=0] at (.25+\x*1.5-1.5,-9) {\y};
\foreach \x/\y in \segcorrs	
	\draw[->,blue] (.25+\x*1.5-1.5,-8.75) -> (.25+\x*1.5-1.5,-8.25);

\end{tikzpicture}
}
\bfcaption{Example for the segment-wise correlations}{We show the signals of two pixels $\fett{x}_{pq}$ (green) and $\fett{x}_{p'q'}$ (red) belonging to the same cell from the Neurofinder training set. The blue lines indicate the boundaries of the ten segments. The results for the segment-wise correlations $c^n$ for $n=1,\dots,10$ are shown beneath the signals. 
The correlation over the whole length of the video between the two pixels is only $c=0.26$. The segment-wise correlations $c^n$, however, are much higher than the overall correlation. This illustrates that the segment-wise computation of the correlations can help identifying pixels belonging to the same cell, especially for cells with very noisy signals and with only sparse activations.  
 \label{fig:disco_seg_corrs}}
\end{figure*}

The segment-wise correlations provide more fine-grained temporal information than the correlations over the whole video. An example illustrating the benefit of segment-wise correlations is shown in figure~\ref{fig:disco_seg_corrs}. 
In this example, we consider two pixels belonging to the same cell. The correlation between the two pixels computed over the whole length of the video is $c=0.26$. Although the pixels belong to the same cell, their overall correlation is rather small as the signals are noisy and activations are sparse in time.  
However, when we split the signals and consider the segment-wise correlations, they are much higher than the overall correlation. The scaled correlation is $c_s=0.69$, thus already much higher than the overall correlation, and the maximum over the segments is even $0.92$. This illustrates that the segment-wise computation of the correlations can help identifying pixels belonging to the same cell. 

Inspired by the idea of the correlation space used in~\citet{spaen_2017}, we compute the correlations not only between a pixel and its direct neighbors, but to a broader neighborhood. For each pixel, we compute the correlation to $C=15$ other pixels with a distance up to three pixels. This extension of the neighborhood empirically showed to provide enough information for the network. Thus, the output of the correlation computations are $N=10$ stacks of size $C\times P\times Q$, where each channel contains the correlations for all $P\times Q$ pixels to one of the 15 considered neighbors.

\subsection{Deep Neural Network}\label{sec:method_dl}
\paragraph{Networks}
The used deep neural network consists of two parts: 
The first CNN with 3D convolutions aggregates the information of the segment-wise correlations. The output of this network is combined with a summary image and passed to the second network, which is a standard 2D U-Net architecture\footnote{We used the U-Net implementation provided in Inferno 0.3.0., see \url{https://github.com/inferno-pytorch/inferno}}~\citep{ronneberger2015u} with depth five. The outputs of this network are the predicted affinities between pixels together with a foreground-background prediction. The details of the network architectures as well as the input and output structures are provided in table~\ref{tab:network} and figure~\ref{fig:disco_model}. The two networks were trained jointly by applying the S{\o}rensen Dice loss~\citep{dice1945measures,sorensen1948method} on the output of the second network. 

\begin{table}[t]
  \caption{DISCo network architecture details. The first network consists of 3D convolutions to aggregate the information from the segment-wise correlations. The output of this network is combined with a summary image and transformed to affinities and a foreground prediction by a U-Net. $N$ is the number of temporal segments, $C$ is the number of neighbors to which the correlations are computed for each of the $P\times Q$ pixels, and $A$ is the number of predicted affinities.}
  \label{tab:network}
  \centering
 \adjustbox{max width=\columnwidth}{
  \begin{tabular}{lllll}
    \toprule
    \multicolumn{5}{c}{Input: segment-wise correlations, tensor of size $C\times N\times P \times Q$} \\
     & Kernel & Channels & Padding  & Activation \\
    3D Convolution & $4\times 3\times 3$ & $2C$ & $0\times 1\times 1$  & ReLU \\
    3D Convolution & $4\times 3\times 3$ & $4C$ & $0\times 1\times 1$  & ReLU \\
    3D Convolution & $4\times 3\times 3$ & $C$ & $0\times 1\times 1$  & ReLU \\
    \multicolumn{5}{c}{Output: aggregated correlations, tensor of size $C\times 1\times P \times Q$} \\
    \midrule
    \multicolumn{5}{c}{Input: aggregated correlations + summary image, } \\
    \multicolumn{5}{c}{tensor of size $(C+1)\times P \times Q$}\\
    U-Net with depth five &&&& Sigmoid\\
    \multicolumn{5}{c}{Output: affinities + foreground prediction, }\\
    \multicolumn{5}{c}{tensor of size $(A+1)\times P \times Q$} \\    
    \bottomrule
  \end{tabular}
}
\end{table}

\paragraph{Training}
The correlations and summary images were normalized channel-wise to zero mean and unit variance. For training we converted the ground truth cell annotations into affinities between pixels and to foreground-background labels. The affinities are computed by first assigning all pixels belonging to a cell with a unique label. For two pixels $i$ and $j$ with assigned labels $L_i$ and $L_j$ the affinity $a_{ij}$ between them is 
\begin{align}
a_{ij} &= \begin{cases} 1 \quad \text{if } L_i = L_j \\ 0 \quad \text{if } L_i \neq L_j \end{cases} \quad .
\end{align}
We applied the channel-wise S{\o}rensen Dice loss~\citep{dice1945measures,sorensen1948method} to all output channels since it has been successfully used to learn affinities~\citep{MWS} and since it can deal with the large class-imbalance that exists between foreground and background in the Neurofinder datasets. 

In order to find suitable hyperparameters for the network training, we split every video of the Neurofinder training set spatially into $\SI{75}{\%}$ training and $\SI{25}{\%}$ validation set. We tested different parameter settings and used the validation loss to determine the best setting. Afterwards, we used this set of hyperparameters to train the networks on the complete videos of the Neurofinder training set. The used hyperparameters can be found in table~\ref{tab:disco_hyperparams}.
For the submission named DISCo we used all Neurofinder datasets, while for DISCos we trained and evaluated on each of the five dataset series individually. 

\begin{table}[t]
\centering
\bfcaption{Hyperparameters used for network training}{\label{tab:disco_hyperparams}}
\adjustbox{max width=.5\textwidth}{
\begin{tabular}{lrrp{2cm}}
\toprule
	& epochs & learning rate  & batch size \\
\midrule
DISCo & 3000 & 0.0001  & 20\\
DISCos & 3000 & 0.0001  & 6 for 00 and 1 otherwise  \\
\bottomrule
\end{tabular}
}
\end{table}

\paragraph{Data Augmentation} 
Since the Neurofinder training data consists of only 19 videos, data augmentation is essential for successful network training. We used data augmentation both in the temporal and the spatial dimensions of the videos as follows: 
Before computing the correlations we performed max-pooling over time on the videos in order to reduce the noisiness of the signals. For training, we varied the temporal length of the max-pooling kernel between three and nine frames. For inference, we fixed the length of the max-pooling kernel to five frames. During train time, we also cut the video randomly into ten segments which were additionally shuffled before computing the correlations. 
In the spatial dimensions of the videos we used random flips and rotations. Additionally, we trained only on random crops of the image plane of size $128\times128$\,pixels. We assured that each crop used for training contained at least one cell. 

\subsection{Instance Segmentation}\label{sec:disco_gasp}
For the final step of extracting the actual cell instances, we use a method called Generalized Algorithm for Signed graph Partitioning (GASP)~\citep{GASP}.
In contrast to the Hochbaum's normalized cut (HNC) model used in HNCcorr~\citep{spaen_2017} and the watershed algorithm used for post-processing in the deep learning models Conv2D~\citep{conv2d} and STNeuroNet~\citep{Ca3dCNN}, GASP neither requires a threshold for stopping the partitioning nor seeds nor a predefined number of clusters.

GASP is designed for the task of partitioning a signed graph $\mathcal{G}=(V,E,W)$ with nodes $V$, edges $E$ and edge weights $W$. In our case, the nodes $V$ correspond to the pixels in the image plane of the calcium imaging video and the structure for the edges is pre-defined as shown in figure~\ref{fig:affinities}. Since cells in calcium imaging videos are usually relatively small, we only connected pixels up to a distance of five pixels to enable the correct separation of small and especially adjacent cells. The weights $w_{ij}\in\mathbb{R}$ for the edges $e_{ij}\in E$ are gained from the predicted affinities $a_{ij}\in\left[0,1\right]$ according to
\begin{align}
w_{ij} &= a_{ij} - 0.5 \quad . 
\end{align}
An edge with high positive edge weight indicates the tendency of the nodes to be merged together in the same cluster, while a strong negative weight corresponds to a strong tendency of the two nodes to be separated.

\begin{figure}[t]
\centering
\adjustbox{max width=.2\textwidth}
{
\begin{tikzpicture}
\foreach \x in {0,...,6}
\foreach \y in {0,...,6}
{
\ifthenelse{\x = 5 \AND \y = 5}{
\node[draw,circle,inner sep=3pt,fill=blue,color=blue] (pixel\x\y) at (\x,-\y) {};
}{
\node[draw,circle,inner sep=3pt,fill=gray!50,color=gray!50] (pixel\x\y) at (\x,-\y) {};
}
}
\draw[color=red, line width=2pt] (pixel55.north) -- (pixel54.south);
\draw[color=red, line width=2pt] (pixel55.west) -- (pixel45.east);
\draw[color=red, line width=2pt] (pixel55.north east) to [out=70,in=-70] (pixel50.south east);
\draw[color=red, line width=2pt] (pixel55.south west) to [out=-170,in=-20] (pixel05.south east);
\draw[color=red, line width=2pt] (pixel55.north west) to [out=120,in=-30] (pixel00.south east);

\end{tikzpicture}
}
\bfcaption{Used edges for GASP}{We show a small grid of pixels. In the signed, edge-weighted graph, each pixel is connected to five pixels in its neighborhood as shown by the red lines for the one exemplary pixel marked in blue. \label{fig:affinities}}
\end{figure}

The GASP algorithm starts with each node in its own cluster and then iteratively merges {\it adjacent} clusters. 
Adjacent means that two clusters $S_u$ and $S_v$ have at least one connecting edge $e_{ij}\in E$ from a node $i$ in cluster $S_u$ to a node $j$ in cluster $S_v$ and that the interaction between the two clusters $\mathcal{W}(S_u,S_v)$ is positive. The interaction or {\it linkage criterion} $\mathcal{W}$ is defined by
\begin{align}\label{eq:disco_gasp_linkage}
\mathcal{W}(S_u,S_v) &= \sum_{e\in E_{uv}} \frac{w_e}{|E_{uv}|} \quad ,
\end{align} 
with $E_{uv}\subset E$ denoting all edges connecting $S_u$ and $S_v$. The {\it average linkage criterion} used in equation~\eqref{eq:disco_gasp_linkage} is just one possible choice among others, e.g.\ sum~\citep{keuper2015efficient,levinkov2017comparative} and absolute maximum linkage~\citep{MWS}. We have chosen the average linkage since it has been shown to be extremely robust and at the same time outperformed other linkage criteria in instance segmentation tasks on biological and street-scene images~\citep{GASP}. The GASP algorithm automatically stops as soon as $\mathcal{W}(S_u,S_v) \leq 0$ for all clusters $S_u$ and $S_v$. 

In order to avoid false merges of cells and background and to remove all background instances from the final result, we exclude all edges to background pixels from the graph before performing the partitioning. The decision whether a pixel is assigned to the background is based on the foreground-background prediction of the network. All pixels for which the background prediction is higher than the value for the foreground / cell prediction are excluded from the graph. This step slightly improved the results and made the instance segmentation step much faster since the graph to partition becomes smaller when excluding all background pixels. As a last step, in order to remove also tiny background segments, we used a simple threshold to exclude all instances with a size smaller than $\SI{25}{pixels}$ from the final result.


\section{Experiments and Results}\label{sec:disco_results}

We trained the networks for DISCo on the publicly available Neurofinder training set with the hyperparameters shown in table~\ref{tab:disco_hyperparams} using the Adam optimizer~\citep{kingma2014adam} and as described in section~\ref{sec:method_dl}. The results are evaluated on the Neurofinder test set with the segmentation quality measured by computing the average F1-score over the nine videos in the test set. 

DISCo significantly outperforms methods which are solely based on summary images like UNet2DS and Conv2D (two-sided Wilcoxon test as described in~\cite{Ca3dCNN}, $p$-value $\leq 0.05$) and also performs much better than those only relying on correlations like HNCcorr, as shown in table~\ref{tab:leaderboard_single} (rows highlighted in gray). 
Furthermore, when training and evaluating individual networks for the five different dataset series 00, 01, 02, 03, and 04 (submission named DISCos), our model outperforms all other methods trained on the Neurofinder datasets (see table~\ref{tab:leaderboard_single}).

\begin{savenotes}
\begin{table*}[t]
\centering
\bfcaption{Neurofinder leaderboard: methods trained only on the Neurofinder datasets}{We show an excerpt of the Neurofinder leaderboard containing the top methods trained on the original Neurofinder training set. Methods using a single model on all five dataset series are highlighted in gray. We show the F1-score for each of the nine test datasets. The sorting is based on the average F1-score over all test datasets ($\varnothing$F1). DISCo outperforms the other methods when applying a single model to all datasets and when training and evaluating individual models on the five dataset series (submission named DISCos). \label{tab:leaderboard_single}}
\adjustbox{max width=\textwidth}{
\begin{tabular}{lrp{.3cm}rrrrrrrrr}
\toprule
Method & $\varnothing$F1 &&\multicolumn{9}{c}{F1-scores on individual test datasets} \\
		& && \multicolumn{2}{c}{00} & \multicolumn{2}{c}{01} & \multicolumn{2}{c}{02} & \multicolumn{1}{c}{03} & \multicolumn{2}{c}{04} \\
\midrule
DISCos\footnote{Network models trained and evaluated on each of the five different groups of datasets individually.\label{fn:fivenets}} 
	& {\bf 0.67} &&0.64 &0.71 &0.61 &0.56 &0.82 &0.77 &0.55 &0.53 &0.82 \\
3dCNN\footref{fn:fivenets} 
	&0.66 && 0.63 &0.72 &0.63 &0.54 & 0.63& 0.56&0.89& 0.55&0.78 \\
\rowcolor{gray!30}
DISCo & {\bf 0.63} &&0.61 &0.73 &0.59 &0.54 &0.60 &0.65 &0.55 &0.55 &0.83 \\
HNCcorr + Conv2D\footnote{Combination of two methods, one applied to datasets 01, 02, 04 and the other to datasets 00, 03\label{fn:twomethods}} 
	& 0.62 &&0.55 & 0.61& 0.53&0.56 &0.75 &0.68 &0.81 &0.38 &0.68 \\
Sourcery\footref{fn:twomethods} 
	& 0.58 &&0.45 &0.53 &0.62 &0.45 &0.72 &0.56 &0.84 &0.39 &0.69 \\
\rowcolor{gray!30}
UNet2DS &0.57&&0.64 &0.70 &0.56 &0.46 &0.49 &0.41 &0.89 &0.33&0.64 \\
Suite2P + Donuts\footref{fn:twomethods}
	&0.55&&0.45 &0.53 &0.49 &0.39 &0.60 &0.52 &0.84 &0.47 & 0.66 \\
$\dots$ && & & & & & & & \\
\rowcolor{gray!30}
HNCcorr & 0.49 &&0.29 &0.33 &0.53 &0.56 &0.75 &0.68 &0.23 &0.38 &0.68 \\
$\dots$ && & & & & & & & \\
\rowcolor{gray!30}
Conv2D & 0.47&& 0.54& 0.61& 0.27&0.27 &0.42 &0.38 &0.84 &0.29 &0.60 \\
\bottomrule
\end{tabular}
}
\end{table*}

\paragraph{Lesion Study}
In order to emphasize how beneficial it is to use both, correlations and summary images, we trained DISCo with different inputs: 
summary images and segment-wise correlations (as proposed in section~\ref{sec:disco_method}); 
summary images and correlations over the whole video length; 
only summary images; and 
only correlations. 

The results in table~\ref{tab:disco_inputs} show that the combination of summary images and correlations over segments clearly outperforms the models only trained on summary images and only on correlations. This holds true when training only a single model on all datasets (see table~\ref{tab:disco_inputs}) but also when training on each of the five dataset series individually (the results for DISCos are shown in the supplementary). Even when combining the results from the models with only summary images and only correlations by using for each dataset series the best result, the proposed approach with the combined input still performs better. This shows that the advantage of the combined input is not only that it can adapt to both kinds of datasets, those with many active cells and those with many inactive cells, but that it also provides more relevant information on all kinds of datasets. 
Comparing the results for the segment-wise correlations and the correlations over the whole video length, the difference is not that big, but still the model using segment-wise correlations performs slightly better. 

\begin{table*}[t]
\centering
\bfcaption{Lesion study: DISCo with different kinds of input}{We show the results on the Neurofinder test set for the DISCo model with different kinds of input. The combination of segment-wise correlations and summary images outperforms the others. 
\label{tab:disco_inputs}}
\adjustbox{max width=\textwidth}{
\begin{tabular}{lrp{.3cm}rrrrrrrrr}
\toprule
Input & $\varnothing$F1 &&\multicolumn{9}{c}{F1-scores on individual test datasets} \\
		& && \multicolumn{2}{c}{00} & \multicolumn{2}{c}{01} & \multicolumn{2}{c}{02} & \multicolumn{1}{c}{03} & \multicolumn{2}{c}{04} \\
\midrule
segment-wise correlations + summary image & {\bf 0.63} &&0.61 &0.73 &0.59 &0.54 &0.60 &0.65 &0.55 &0.55 &0.83 \\
correlations (whole video) + summary image &0.62 && 0.47 & 0.70 & 0.62 & 0.53 & 0.74 & 0.67 & 0.55 & 0.50 & 0.83 \\
only correlations &0.50 && 0.32 & 0.42 & 0.45 & 0.41 & 0.68 & 0.62 & 0.36 & 0.50 & 0.77\\
only summary image &0.49 && 0.51 & 0.59 & 0.51 & 0.38 & 0.51 & 0.37 & 0.54 & 0.31 & 0.65 \\
best of only correlations and only summary image & 0.57 &&  0.51 & 0.59 & 0.51 & 0.38 & 0.68 & 0.62 & 0.54 & 0.50 & 0.77\\
\bottomrule
\end{tabular}
}
\end{table*}
 
 \end{savenotes}

An alternative to using a CNN to aggregate the information from the ten segments into a single input for the second network would be to use the scaled correlation~\citep{nikolic2012scaled}. However, in table~\ref{tab:disco_temp_comp} we show that using only the mean over the segments performs much worse than using the aggregation network. This is also the case when taking the maximum over the segments instead of the mean. Only when we use a combination of several statistics, namely maximum, minimum, mean, standard deviation, and sum, the model performs almost as good as the model using the CNN. 

Using 3D convolutions in the aggregation network might at first seem counter-intuitive, since the final segmentation and hence also the output of the aggregation network should be invariant under permutations of the $N$ segments of the video. For this reason, we also tested an aggregation network with 2D convolutions in which the temporal information is aggregated through max-pooling over the $N$ segments at the end. Although we slightly increased the number of layers and used kernels for the 2D network compared to the network with 3D convolutions, the 2D network did not reach the same performance as the 3D model (see table~\ref{tab:disco_temp_comp}). We assume that one reason is that the temporal information can be aggregated in a more complex way by using the 3D convolutions. Although the invariance of the network output under permutations of the segments is neither directly enforced nor guarantied, we assume that the needed permutation invariance is approximately realized through randomly shuffling the segments during train time and by using a large enough number of convolution kernels. 

\begin{table*}[t]
\centering
\bfcaption{Lesion study: DISCo with different aggregation models}{We compare the results of DISCo on the Neurofinder test set using different statistics over the ten segments with the use of a CNN. The model using the CNN outperforms the alternatives. 
\label{tab:disco_temp_comp}}
\adjustbox{max width=\textwidth}{
\begin{tabular}{lrp{.3cm}rrrrrrrrr}
\toprule
Aggregation from segments & $\varnothing$F1 &&\multicolumn{9}{c}{F1-scores on individual test datasets} \\
to single input		& && \multicolumn{2}{c}{00} & \multicolumn{2}{c}{01} & \multicolumn{2}{c}{02} & \multicolumn{1}{c}{03} & \multicolumn{2}{c}{04} \\
\midrule
3D CNN & {\bf 0.63} &&0.61 &0.73 &0.59 &0.54 &0.60 &0.65 &0.55 &0.55 &0.83 \\
2D CNN &  0.61 && 0.55 & 0.69 & 0.58 & 0.50 & 0.66 & 0.61 & 0.55 & 0.54 & 0.83\\
mean (scaled correlation) & 0.61 && 0.52 & 0.72 & 0.58 & 0.51 & 0.68 & 0.63 & 0.55 & 0.50 & 0.78\\
max & 0.60 && 0.58 & 0.71 & 0.55 & 0.51 & 0.60 & 0.58 & 0.56 & 0.48 & 0.82\\
max, min, mean, std, and sum & 0.62 && 0.49 & 0.72 & 0.60 & 0.54 & 0.69 & 0.70 & 0.54 & 0.47 & 0.79\\
\bottomrule
\end{tabular}
}
\end{table*}

\paragraph{Training with only one Video}
In practice, the calcium imaging data to be segmented is expected to differ more or less drastically from the data provided in the Neurofinder challenge. One way to approach this problem is to train the model on as many datasets as possible and hope that the model is able to generalize well, as e.g.\ done in STNeuroNet. However, given the huge variety of calcium indicators, recording setups, and brain regions to observe, we doubt that a single model is able to perform well under {\it all} conditions. Moreover, depending on the research question the model might be expected to detect only active cells or active {\it and} inactive cells.\footnote{See e.g.\ the comments on the labeling criteria used in the Neurofinder dataset series 04 at \url{https://github.com/codeneuro/neurofinder/issues/25}} 

For this reason, we decided to test our model in a different direction, namely when being trained only on a single video. 
We think it is a realistic scenario that for a new recording setup, which differs too much from previous datasets to use an already trained model, a neuroscience expert could manually label {\it one} video. This one video would then be used to train a model for segmenting the wanted cells under the given recording conditions. Afterwards the trained model could be applied to other videos recorded under similar conditions. 

In order to find out how well DISCo would perform in such a scenario, we trained DISCo always on only a single video and then evaluated the performance of the trained model on all other videos from the same dataset series, including the remaining videos from the training set as well as the videos from the test set. The results are shown in detail in the supplementary. A first finding of these results is that the performance of the model depends on the video used for training. In all cases, the performance of the model on the other training datasets is much better than the performance on the test set. In order to find out what the reason for this inequality between training and test set could be and why some datasets seem to be better suited for training the model than others, it would be necessary to let a neuroscience expert analyze the different datasets and the ground truth annotations used for training and testing. 

The average performance of DISCo trained on only one video is still quite good. An average score of $0.58$ on the test sets is comparable to the performance of Sourcery. 

\paragraph{Precision and Recall}
Considering the average F1-score, DISCo shows great performance on the Neurofinder datasets. When investigating the precision and recall achieved on the different test datasets, however, we notice that on some datasets DISCo suffers from a relatively low recall compared to its precision, while on other datasets the precision is the limiting factor for the overall performance. An explanation why DISCo has problems to achieve a good recall especially on the datasets from series 00, while it struggles with the precision mainly on datasets from series 01, has not been found yet. Further analyses of the results and of the properties of the different datasets are required to find the reason for these different behaviors of the model.


\section{Summary and Conclusion}\label{sec:disco_conclusion}
We have presented a new approach for the cell segmentation in calcium imaging videos. We use a deep learning model, but in contrast to previous work we neither only rely on purely shape-based summary images as input nor use computationally expensive 3D CNNs directly on the calcium imaging videos. Instead, we propose a fast and computationally efficient framework that achieves top scores on the Neurofinder benchmark. As input to our network model we use a combination of correlations between pixels and summary images, which allows us to detect active cells as well as cells with weak or almost constant signals, and the presented scheme of computing the correlations segment-wise provides more fine-grained temporal information than the correlation over the whole length of the video. 

Another novelty compared to other methods for cell segmentation in calcium imaging videos is that the used deep learning model does not only provide a foreground-background prediction, but additionally predicts affinities between pixels. This allows us to directly apply an instance segmentation method on the network output to extract the individual cells.  

In future work it would be interesting to further analyze the datasets and the results in order to find out why on some datasets the overall performance of the method is limited by the recall while on others by the precision. 

{\small
\bibliographystyle{unsrtnat}
\bibliography{references}
}


\section*{Supplementary Material}

\begin{table*}
\centering
\bfcaption{Results with DISCo for different kinds of input}{We show the results on the Neurofinder test set for the DISCo model with different kinds of input. Both, when training only a single model on all datasets (top) and when training individual network models on the five dataset series (bottom), the combination of segment-wise correlations and summary images clearly outperforms the others. 
\label{tab:disco_inputs}}
\adjustbox{max width=\textwidth}{
\begin{tabular}{lrp{.3cm}rrrrrrrrr}
\toprule
Input & $\varnothing$F1 &&\multicolumn{9}{c}{F1-scores on individual test datasets} \\
		& && \multicolumn{2}{c}{00} & \multicolumn{2}{c}{01} & \multicolumn{2}{c}{02} & \multicolumn{1}{c}{03} & \multicolumn{2}{c}{04} \\
\midrule
{\bf DISCo:} & \\
correlations (segments)  & {\bf 0.63} &&0.61 &0.73 &0.59 &0.54 &0.60 &0.65 &0.55 &0.55 &0.83 \\
\hspace{.3cm}+ summary images & \\
correlations (whole video)  &0.62 && 0.47 & 0.70 & 0.62 & 0.53 & 0.74 & 0.67 & 0.55 & 0.50 & 0.83 \\
\hspace{.3cm}+ summary images & \\
only correlations (segments) &0.50 && 0.32 & 0.42 & 0.45 & 0.41 & 0.68 & 0.62 & 0.36 & 0.50 & 0.77\\
only summary image &0.49 && 0.51 & 0.59 & 0.51 & 0.38 & 0.51 & 0.37 & 0.54 & 0.31 & 0.65 \\
best of only correlations and & 0.57 &&  0.51 & 0.59 & 0.51 & 0.38 & 0.68 & 0.62 & 0.54 & 0.50 & 0.77\\
\hspace{.3cm} only summary images & \\
\midrule
{\bf DISCos:} & \\
correlations (segments)  &  {\bf 0.67} &&0.64 &0.71 &0.61 &0.56 &0.82 &0.77 &0.55 &0.53 &0.82 \\
\hspace{.3cm}+ summary images & \\
correlations (whole video) &0.66 && 0.65 & 0.72 & 0.63 & 0.54 & 0.79 & 0.76 & 0.52 & 0.53 & 0.84 \\
\hspace{.3cm}+ summary images & \\
only correlations (segments) &0.55 && 0.35 & 0.44 & 0.46 & 0.43 & 0.78 & 0.73 & 0.51 &0.47 & 0.80  \\
only summary image & 0.58 && 0.66 & 0.69 & 0.58 & 0.49 & 0.66 & 0.63 & 0.54 & 0.35 & 0.68 \\
best of only correlations and & 0.64 && 0.66 & 0.69 & 0.58 & 0.49 & 0.78 & 0.73 & 0.54 & 0.47 & 0.80 \\
\hspace{.3cm}only summary images & \\
\bottomrule
\end{tabular}
}
\end{table*}

\begin{table*}
\centering
\bfcaption{Results with DISCo for different temporal compression models}{We show the results on the Neurofinder test set for the DISCo model using different ways of temporal compression for the segment-wise correlations. We compare different statistics over the ten segments with the use of a temporal compression network and the computation of the correlations over the whole video length. Both, when training only a single model on all datasets (top) and when training individual network models on the five dataset series (bottom), the model using the temporal compression network outperforms the alternatives. 
\label{tab:disco_temp_comp}}
\adjustbox{max width=\textwidth}{
\begin{tabular}{lrp{.3cm}rrrrrrrrr}
\toprule
Conversion from segments & $\varnothing$F1 &&\multicolumn{9}{c}{F1-scores on individual test datasets} \\
to single input		& && \multicolumn{2}{c}{00} & \multicolumn{2}{c}{01} & \multicolumn{2}{c}{02} & \multicolumn{1}{c}{03} & \multicolumn{2}{c}{04} \\
\midrule
{\bf DISCo:} & \\
temporal compression network & {\bf 0.63} &&0.61 &0.73 &0.59 &0.54 &0.60 &0.65 &0.55 &0.55 &0.83 \\
only mean & 0.61 && 0.52 & 0.72 & 0.58 & 0.51 & 0.68 & 0.63 & 0.55 & 0.50 & 0.78\\
only max & 0.60 && 0.58 & 0.71 & 0.55 & 0.51 & 0.60 & 0.58 & 0.56 & 0.48 & 0.82\\
max + min + mean + std + sum & 0.62 && 0.49 & 0.72 & 0.60 & 0.54 & 0.69 & 0.70 & 0.54 & 0.47 & 0.79\\
correlations over whole video  &0.62 && 0.47 & 0.70 & 0.62 & 0.53 & 0.74 & 0.67 & 0.55 & 0.50 & 0.83 \\
\midrule
{\bf DISCos:} & \\
temporal compression network  &  {\bf 0.67} &&0.64 &0.71 &0.61 &0.56 &0.82 &0.77 &0.55 &0.53 &0.82 \\
only mean & 0.64 && 0.55 & 0.67 & 0.63 & 0.55 & 0.78 & 0.69 & 0.54 & 0.55 & 0.83\\
only max & 0.64 && 0.57 & 0.68 & 0.59 & 0.53 & 0,78 & 0.75 & 0.54 & 0.54 & 0.82\\
max + min + mean + std + sum & 0.66 && 0.61 & 0.70 & 0.64 & 0.54 & 0.79 & 0.75 & 0.55 & 0.54 & 0.81\\
correlations over whole video &0.66 && 0.65 & 0.72 & 0.63 & 0.54 & 0.79 & 0.76 & 0.52 & 0.53 & 0.84 \\
\bottomrule
\end{tabular}
}
\end{table*}

\begin{table*}
\centering
\bfcaption{Results for DISCo when being trained only on a single video}{We trained DISCo on each of the 19 training videos individually and evaluated its performance on the remaining training and test datasets of the same dataset series. We show the average F1-scores ($\varnothing$F1) with corresponding standard deviation. In the bottom line we also show the overall mean with standard deviation for the results on the training sets, the test sets and the combination of training and test sets. \label{tab:disco_onevideo}}
\adjustbox{max width = \columnwidth}{
\begin{tabular}{lrrr}
\toprule
trained on & $\varnothing$F1 & $\varnothing$F1 & $\varnothing$F1  \\
		& train set & test set & train + test set \\
\midrule
00.00 & $0.51\pm 0.09$ & \qquad $0.44\pm 0.03$ & $0.50\pm 0.08$  \\
00.01 & $0.72\pm 0.12$ & $0.52\pm 0.04$ & $0.69\pm 0.13$ \\
00.02 & $0.78\pm 0.12$ & $0.53\pm 0.03$ & $0.74\pm 0.14$ \\
00.03 & $0.60\pm 0.13$ & $0.50\pm 0.03$ & $0.59\pm 0.13$ \\
00.04 & $0.86\pm 0.07$ & $0.62\pm 0.02$ & $0.82\pm 0.11$ \\
00.05 & $0.86\pm 0.07$ & $0.65\pm 0.04$ & $0.82\pm 0.10$ \\
00.06 & $0.76\pm 0.14$ & $0.54\pm 0.01$ & $0.73\pm 0.16$ \\
00.07 & $0.77\pm 0.12$ & $0.59\pm 0.04$ & $0.74\pm 0.13$ \\
00.08 & $0.77\pm 0.12$ & $0.55\pm 0.08$ & $0.74\pm 0.14$ \\
00.09 & $0.91\pm 0.07$ & $0.66\pm 0.03$ & $0.87\pm 0.11$ \\
00.10 & $0.86\pm 0.08$ & $0.58\pm 0.03$ & $0.82\pm 0.13$ \\
00.11 & $0.73\pm 0.14$ & $0.51\pm 0.05$ & $0.67\pm 0.15$ \\
01.00 & 0.86 & $0.58\pm 0.07$ & $0.67\pm 0.14$ \\
01.01 & 0.80 & $0.57\pm 0.03$ & $0.64\pm 0.11$ \\
02.00 & 0.89 & $0.78\pm 0.01$ & $0.81\pm 0.06$ \\
02.01 & 0.99 & $0.67\pm 0.02$ & $0.78\pm 0.15$ \\
03.00 & -- & 0.56 & 0.56\\
04.00 & 0.88 & $0.54\pm 0.03$ & $0.65\pm 0.16$ \\
04.01 & 0.80 & $0.62\pm 0.19$ & $0.68\pm 0.18$ \\
\midrule
$\varnothing$ & $0.77\pm 0.15$ & $0.58\pm 0.10$ & $0.73\pm 0.16$ \\
\bottomrule 
\end{tabular}
}
\end{table*}

\end{document}